\documentclass[%
reprint,
superscriptaddress,
amsmath,
amssymb,
aps,
pra,
]{revtex4-2}

\usepackage{graphicx}
\usepackage{dcolumn}
\usepackage{bm}
\usepackage{bbold}
\usepackage[hypertexnames=false]{hyperref}
\hypersetup{
	colorlinks = true,
	allcolors = blue,
}

\newcommand{\bL}{{\bm L}}

\begin{document}
\title{Bose-Einstein condensate in a shell: nonlinear modes and symmetry breaking}
\author{Alexander Wolf}
\email{a.wolf@dlr.de}
\affiliation{German Aerospace Center (DLR), Institute of Quantum Technologies, 89081 Ulm, Germany}
\affiliation{Institut f\"ur Quantenphysik and Center for Integrated Quantum Science and Technology (IQST), Universit\"at Ulm, 89069 Ulm, Germany}
\author{Vladimir V. Konotop}
\affiliation{Departamento de Física and Centro de Física Teórica e Computacional, Faculdade de Ciências, Universidade de Lisboa, Lisboa, Portugal}
\author{Maxim A. Efremov}
\affiliation{German Aerospace Center (DLR), Institute of Quantum Technologies, 89081 Ulm, Germany}
\affiliation{Institut f\"ur Quantenphysik and Center for Integrated Quantum Science and Technology (IQST), Universit\"at Ulm, 89069 Ulm, Germany}
\date{\today}
\begin{abstract}
The state of a quantum particle confined to a spherical surface is commonly described using spherical harmonics, which play a crucial role in solving many scientific problems featuring spherical symmetry. A mean-field description of a system of many interacting quantum particles confined to a spherical shell naturally leads to nonlinear generalizations of spherical harmonics. In this study, we analyze the emergence of such nonlinear modes in a shell-shaped Bose-Einstein condensate with both repulsive and attractive two-particle interactions. We find several classes of nonlinear modes. The first kind is an analytical continuation of a single spherical harmonic or their superposition for nonvanishing atom-atom interaction. Some of these solutions have planar counterparts in the form of Townes solitons and Townes vortices, which either appear stable even in the case of attractive interactions or can be stabilized by three-particle interactions. The second kind originate from symmetry-breaking bifurcations in the case of a repulsive interaction, and therefore have no linear counterparts. Moreover, we demonstrate the existence of a breathing dynamics that oscillates primarily between an unstable soliton and a stable vortex necklace state. Our findings show that the physics of nonlinear matter waves on curved geometries differs substantially from that in planar systems, suggesting the feasibility of observing these unconventional wave patterns experimentally.
\end{abstract}
\maketitle
\section{Introduction}
Many phenomena in nature are determined by underlying geometry, in particular, by the number of dimensions and space topology. Here, the term "space" may refer to configuration (coordinate) space, momentum space, or more generally to an abstract parameter space. For example, in the random walk~\cite{Feller.1957}, which is a fundamental process in physics, chemistry, biology, engineering, economics, sociology, etc., the probability to return to the origin is unity in one and two dimensions but smaller than 0.35 for dimensions larger than three. In addition, Bose-Einstein condensation~\cite{Dalfovo.1999} occurs only in three spatial dimensions and does not exist in one and two. Moreover, the Huygens-Fresnel principle~\cite{Born.1980} describing the wavefront dynamics of electromagnetic waves is only valid in odd spatial dimensions. The topology of space brings constants of motion into play and results in concepts such as geometrical phases~\cite{Cohen.2019,Bohm.2003}, as well as topologically- and symmetry-protected states~\cite{Qi.2011,Chiu.2016,Ozawa.2019}.

The recent progress in realizing curved spaces in a lab, e.g. quantum bubbles or spherical shells~\cite{Lundblad.2023,Tononi.2023,Dubessy.2025}, planar rings~\cite{Amico.2022}, and Möbius strips~\cite{Barredo.2018}, offers a new avenue to explore the interplay between physics and geometry. Since the surface of a sphere represents a two-dimensional curved space with constant Gaussian curvature, spherical shells have become a popular platform for reexamining Bose-Einstein condensation~\cite{Tononi.2019,Moller.2020}, collective excitations~\cite{Wolf.2022,Lannert.2007,Sun.2018}, vortices~\cite{Turner.2010,Padavic.2020,Bereta.2021}, solitons, and confinement-induced resonances~\cite{Carmesin.2025}.

Currently, there are two working schemes~\cite{Lundblad.2023} for producing atomic shells in laboratory environments.  The first employs radio-frequency dressing~\cite{Zobay.2001, Zobay.2004,Garraway.2016} in a microgravity environment, resulting in ellipsoidal shells~\cite{Carollo.2022,Guo.2022b}. The second scheme involves optically confined mixtures of two BECs and either uses a magic laser wavelength~\cite{Jia.2022} or microgravity~\cite{Wolf.2022}. Combined with Feshbach resonances~\cite{Chin.2010,Timmermans.1999} to tune the interspecies interaction, one of the mixture components forms a spherical shell around the other component. 

In this paper, we report on the emergence of stable nonlinear modes in a Bose-Einstein condensate (BEC) with both repulsive and attractive two-body interaction confined within a spherical shell. We identify the structure and origin of the stability of each found mode and analyze the role of (i) the amplitude and sign of the interparticle interaction and (ii) the compactness and curvature of the spherical shell.
\section{Angular GPE}
The macroscopic wavefunction $\Psi(r,\theta,\varphi,t)$, where $r$ is the radial component, $\theta$ is the polar angle, and $\varphi$ is the azimuthal angle, of a BEC trapped tightly to a spherical surface can be approximately factorized according to $\Psi(r,\theta,\varphi,t) \approx \exp\left(-\mathrm{i} \varepsilon t / \hbar\right) \psi(\theta,\varphi,t) u(r) / r$. Here, the radial wavefunction $u(r)$ satisfies
\begin{equation}
    -\frac{\hbar^2}{2 M} \partial_r^2 u + V(r) u = \varepsilon u
\end{equation}
with normalization $\int_0^\infty \mathrm{d} r \, u^2(r) = 1$. The trapping potential $V(r)$ significantly differs from zero in a narrow vicinity of the spherical surface having the radius $R$. The wavefunction $\psi(\theta,\varphi,t)$ describing the angular dependence, satisfies the reduced Gross-Pitaevskii equation (GPE)
\begin{equation}\label{eq:GPE}
    \mathrm{i} \partial_t \psi = \frac{1}{2} \bL^2\psi + G |\psi|^2\psi,
\end{equation}
where $\bL^2 = -\Delta_\theta - \sin^{-2}(\theta) \partial_\varphi^2$ with $\Delta_\theta = \partial_\theta^2 + \cot(\theta) \partial_\theta$ being the squared angular momentum operator (see e.g.~\cite{Landau.2003} and the supplemental material). In Eq.~\eqref{eq:GPE} time is measured in units of $\tau \equiv  I / \hbar$ where
\begin{equation}\label{eq:inertia}
    I = M \left(\int_{0}^\infty \mathrm{d}r \ \frac{u^2}{r^2}\right)^{-1}
\end{equation}
is the moment of inertia and $M$ is the mass of the atoms in the BEC. Here, the angular wavefunction is normalized to one $\int \sin(\theta) \mathrm{d}\theta \mathrm{d}\varphi \ |\psi|^2 = 1$. With this choice of normalization we have the dimensionless effective nonlinearity $G$ given by
\begin{equation}\label{eq:G}
    G = \frac{4 \pi a N}{\sigma}, \quad \frac{1}{\sigma} = \frac{\int_{0}^\infty \mathrm{d}r \ u^4 / r^2}{\int_{0}^\infty \mathrm{d}r \ u^2 / r^2}
\end{equation}
where $a$ is the three-dimensional $s$-wave scattering length, $N$ is the number of atoms, and $\sigma$ is the form factor. In the tight-confinement regime, where $u(r)$ is strongly localized in the spherical shell with the width $\Delta R$, we can estimate the moment of inertia~\eqref{eq:inertia} to be $I \approx M R^2$, as well as the nonlinear parameter~\eqref{eq:G} $G\approx 4\pi a N/\Delta R$.

We are interested in stationary solutions of the angular GPE~\eqref{eq:GPE} and use the ansatz $\psi(\theta,\varphi,t) = \Phi(\theta,\varphi) \mathrm{e}^{- \mathrm{i} \mu t}$, where the function $\Phi(\theta,\varphi)$ and corresponding chemical potential satisfy the reduced stationary angular GPE
\begin{equation}\label{eq:eigenvalue}
    \mu \Phi = \frac{1}{2} \bL^2 \Phi + G |\Phi|^2 \Phi \,.
\end{equation}
At $G = 0$ Eq.~\eqref{eq:eigenvalue} is a linear eigenvalue problem, $\tilde\mu \tilde\Phi = \frac{1}{2} \bL^2 \tilde\Phi$ (hereafter tilde stands for the linear limit) whose solutions are the well-known spherical harmonics~\cite{Landau.2003} $\tilde{\Phi}_{lm} = Y_{lm}(\theta,\varphi) \propto P_l^m(\cos\theta) \mathrm{e}^{\mathrm{i} m \varphi}$ with $l = 0,1,\ldots$, $m = -l,\ldots,l$, the associated Legendre polynomials $P_l^m(\cos\theta)$, and eigenvalues $\tilde{\mu}_l = l (l + 1) / 2$.  The spectrum is degenerate in the polar (magnetic) quantum number $m$ and depends only on the azimuthal quantum number $l$. Hence, superpositions of spherical harmonics with the same $l$ but different $m$ are stationary states as well.

Below we find that the individual spherical harmonics and certain superpositions allow analytic continuation to the nonlinear regime governed by the parameter $G \neq 0$. They can be identified via the functional dependence $G(\mu)$ and we refer to them according to the $l$ and $m$ numbers that correspond to their linear limit $G(\tilde{\mu}_l) = 0$. 

We begin by considering the subset of solutions with embedded vorticity $m = 0, \pm 1, \ldots$, which are analytic continuations of the linear spherical harmonics and are denoted by $\Phi_{lm}(\theta,\varphi) = \phi_{lm} (\theta) \mathrm{e}^{\mathrm{i} m \varphi}$. Thus, $\Phi_{lm}(\theta,\varphi) \to Y_{lm}(\theta,\varphi)$ and $\phi_{lm} (\theta) \to P_l^m(\cos\theta)$ at $G \to 0$. The $\phi_{lm}(\theta)$ can always be chosen real (see supplemental material for a proof). Afterwards we discuss more complicated states which arise either due to symmetry-breaking bifurcations or due to continuations of superpositions.

Stability of each stationary state is investigated by solving the respective Bogoliubov-de Gennes (BdG) equations [Eq.~\eqref{eq:bdgMatrix} in the Appendix] yielding the linear excitation spectrum $\omega$ of small perturbations around the state, discussed in detail in the Appendix. If any eigenvector displays a complex eigenvalue, the state is unstable, otherwise we consider it stable. For solutions with embedded vorticity, the BdG spectrum separates into subsets as well, which we label using the azimuthal $\mathfrak{l}$ and polar $\mathfrak{m}$ quantum numbers.
\section{Nonlinear harmonics without vorticity}
\begin{figure*}[tbhp]
	\centering
    \includegraphics[width=452.34167pt,height=522.27643pt]{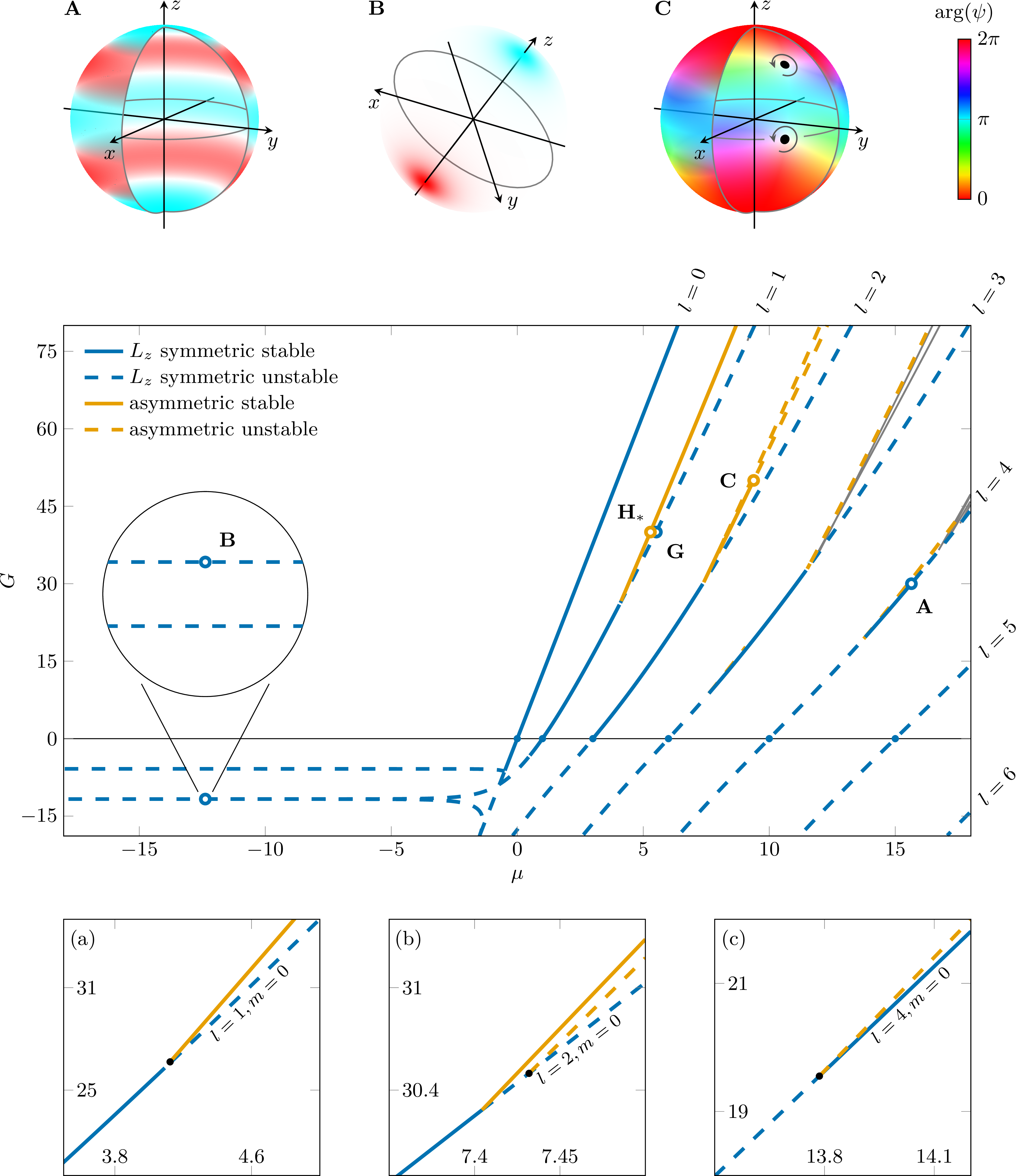}
    \caption{\label{fig:m=0}
        \textbf{Nonlinear continuation of spherical harmonics without embedded vorticity.} Central panel shows families of nonlinear modes on the $(\mu,G)$ diagram. Intervals of linearly stable and unstable solutions are displayed as solid and dashed lines, respectively. Color indicates whether the solution is $L_z$ symmetric (blue) or asymmetric (orange). Solution families bifurcating from the linear limit $G = 0$, blue dots, are marked by the corresponding $l$ quantum number attached outside the frame. Gray lines identify more unstable families resulting from bifurcations. Three states discussed in the text are illustrated to the top with the color representing the phase of the wavefunction $\psi$ and opacity the modulus. Bottom panels show three different scenarios of symmetry-breaking bifurcations indicated by black dots. The lower bifurcation in panel (b) is of the type shown in (a).
    }
\end{figure*}

We begin with states that have zero embedded vorticity $m = 0$. They posses an $L_z$ axial symmetry because they lack a $\varphi$-dependence and do not change when rotated around the $z$-axis. The numerical results are summarized in  Fig.~\ref{fig:m=0}, showing that all linear states allow continuation to the nonlinear domain.

The fundamental state is the homogeneous solution $\Phi_{00} = 1 / \sqrt{4 \pi}$ which preserves its shape in the nonlinear case and is defined by the linear dependence $G = 4 \pi \mu$. This is the shell-shaped BEC with a constant atomic density realized in recent experiments~\cite{Carollo.2022,Jia.2022}. While it is an obvious stationary solution, is has nontrivial stability properties. 

Indeed the BdG equations yield the excitation spectrum
$
\omega_{\mathfrak{l}} = \pm (1/2)\sqrt{\mathfrak{l}(\mathfrak{l}+1)\bigl[\mathfrak{l}(\mathfrak{l}+1) + G/\pi\bigr]},
$
which is $(2\mathfrak{l}+1)$-fold degenerate with respect to the quantum number $\mathfrak{m} = -\mathfrak{l},\ldots,\mathfrak{l}$ for $\mathfrak{l}=0,1,\ldots$. Consequently, although the ground state is (as expected) dynamically stable for repulsive interactions $G>0$ (corresponding to $\mu>0$), it can also remain stable in the presence of attractive interactions $G<0$, since $\omega_{\mathfrak{l}}$ remains real in the interval $-2\pi < G < 0$.

More generally, the numerical analysis shows that the states $\Phi_{l0}$ with $l > 0$ have extensive regions of stability in the nonlinear repulsive regime as well (we have checked up to $l = 4$). These nonlinear Legendre modes are characterized by $l$ stripes at different latitudes as displayed in Fig.~\ref{fig:m=0} for $l = 4$ (state \textbf{A}). We also note that for $G > 0$ the solution $\Phi_{10}$  has been investigated in Ref.~\cite{Sallatti.2026}. 

Turning to attractive interaction $G < 0$ we observe in Fig.~\ref{fig:m=0} that the nonlinear continuation of the two lowest linear solutions with  $l = 0$ and $l=1$, have a finite stability region. This is due to the compactness of the spherical surface, resulting in a discrete spectrum. Indeed, since  the BdG equation obeys pseudo-Hermicity, its complex eigenvalues can emerge only in complex-conjugate pairs after "collision" of two real eigenvalues. Since the linear solution is stable and the distance between neighboring eigenstates is finite, for them to collide requires non-negligible $G$, i.e., above a certain threshold (see e.g. \cite{Caliceti.2005}). For example, in the BdG spectrum $\omega_{\mathfrak{l}}$ of the homogeneous state $\Phi_{00}$ the first collision of neighboring eigenvalues occurs at $\omega_{\mathfrak{l}} = 0$ when $G = -2\pi$ and is achieved for $\mathfrak{l}=1$ ($\mathfrak{l} = 0$ is the Goldstone mode of the $U(1)$-symmetry breaking).

Beyond the instability onset discussed above, the family of nonlinear solutions may also experience bifurcations, which is another scenario of the loss of stability. One illustrative example is the case of  homogeneous state $\Phi_{00}$, which undergoes symmetry-breaking bifurcations of the form $\mathrm{SO}(3)\to \mathrm{SO}(2)$, such that only axial ($L_z$) symmetry is retained. This transition is evident in Fig.~\ref{fig:m=0} at $G=-2\pi$, where the homogeneous state loses stability.

The state that bifurcates away quickly approaches the value of $G \approx -5.85$ from below and becomes almost a horizontal line in the $G$-$\mu$-diagram, Fig.~\ref{fig:m=0}. This behavior is a hallmark of the Townes soliton family~\cite{Chiao.1964} which are well-known bright solitonic solutions of the planar 2D nonlinear Schrödinger equation~\cite{Malkin.1991}. That equation is the flat counterpart to the angular GPE~\eqref{eq:GPE} and possesses a rescaling symmetry: it transforms a solution to any length scale $\lambda$ by the change of variables $\mathbf{x} = \mathbf{r} / \lambda$ with the new chemical potential $\mu_\lambda = \lambda^2 \mu$. The planar Townes solitons therefore appear in the $G$-$\mu$-diagram as perfect horizontal lines at negative $\mu$ and discrete values of $G$.

The angular GPE~\eqref{eq:GPE} has no scaling symmetry. Unlike the planar case, the fixed curvature of the sphere introduces an intrinsic length scale, so the Laplace–Beltrami operator $\Delta_{\mathbb{S}^2} = -\bL^2$ is invariant only under the rotational isometries $\text{SO(3)}$ and not under dilations. However, when solutions become sufficiently localized the effect of the curvature diminishes and the rescaling symmetry is asymptotically recovered. Consider, for example, a solution strongly localized at the north pole of an arbitrarily fixed coordinate system. Then, for $|\theta| \ll 1$  the Laplace–Beltrami operator can be approximated as $\Delta_{\mathbb{S}^2} \approx \partial_\theta^2 + \theta^{-1} \partial_\theta + \theta^{-2} \partial_\varphi^2$ which is the planar 2D Laplacian expressed in terms of polar coordinates. This explains why the respective solutions on the sphere, when becoming sufficiently narrow, approach planar Townes solitons and $G \to - G_\text{TS}$ where $G_\text{TS} \approx 5.85$ is the norm of the standard Townes soliton.

More general solutions for attractive interactions include narrow humps located at opposite poles which can be interpreted as a superposition of two Townes solitons. An example within the subset of solutions with zero embedded vorticity is the $l = 1$ state which approaches $G = -2 \times G_\text{TS} \approx -11.7$ from above and approximates Townes solitons of opposite signs centered at different poles (Fig.~\ref{fig:m=0}, state \textbf{B}). However, a state with the solitons localized at the poles but now having the same sign bifurcates from the homogeneous state at $(G,\mu) = (-6 \pi, -3 / 2)$ in Fig.~\ref{fig:m=0} (central panel) and approaches $-2 \times G_{\rm TS}$ from below.

It turns out that the Townes-like solutions are unstable similar to the planar case, as Fig.~\ref{fig:m=0} shows. However, we have found that a small repulsive higher order nonlinearity $\delta |\psi|^4 \psi$ with $\delta > 0$ stabilizes a single polar soliton similarly to the known case of the planar geometry~\cite{Malkin.1991}. The two-hump state \textbf{B} in Fig.~\ref{fig:m=0} where the humps are out-of-phase is also stabilized unlike the in-phase two-hump mode.
\section{Symmetry-breaking bifurcations}
In addition to the embedded vorticity structures arising in the nonlinear continuation of the linear spherical harmonics $Y_{lm}$ with $m\geq 1$ (discussed below), an alternative mechanism leading to solutions with nontrivial $\varphi$-dependence are bifurcations that break the $L_z$ symmetry of a given state. We investigate this scenario for the states without vorticity $m = 0$ shown in Fig.~\ref{fig:m=0} in the regime of positive scattering length $G > 0$. Three cases are displayed in the bottom panels and indicated by the black dot: (a) a stable symmetric solution becomes unstable and the new family picks up the stability, (b) all solutions are unstable before and after the bifurcation, and (c) an unstable symmetric solution becomes stable as the new symmetry-broken family takes the instability with it.

Case (a) illustrates a supercritical pitchfork bifurcation resulting in a family of stable symmetry-broken solutions (solid orange lines in Fig.~\ref{fig:m=0}). Such bifurcations are found  for the families without embedded vorticity having $l = 1$, $2$, and $3$. They are identified by the collisions of BdG eigenvalues at $\omega = 0$ within the $|\mathfrak{m}| = 2$, $3$, and $4$ subspaces, respectively (see supplemental material). The symmetry-broken solutions feature $2 |\mathfrak{m}|$ vortices at each soliton stripe of the symmetric state for $2 l |\mathfrak{m}|$ vortices in total. Only the family bifurcated from the state $\Phi_{10}$ at $G \approx 26.3$, Fig.~\ref{fig:breather} state \textbf{H}, has an extended region of stability. The families bifurcated from $\Phi_{20}$ at $G \approx 30.28$ and $\Phi_{30}$ at $G \approx 32.89$ become unstable slightly beyond the bifurcation.

The new symmetry-broken family resulting from the bifurcation in case (b) is unstable. However, it may become stable at some point beyond the bifurcation. An example for this is the family bifurcating from $\Phi_{20}$ at $G \approx 30.5$ which is initially unstable before becoming stable in the interval $36 \lesssim G \lesssim 51$, i.e. outside the region shown in Fig.~\ref{fig:m=0} panel (b). Solutions in this family have two pairs of vortices and two pairs of antivortices where each pair has a vortex above and below the equator, depicted as state \textbf{C} in Fig.~\ref{fig:m=0} with one pair visible.

Finally, case (c) illustrates a subcritical pitchfork bifurcation resulting in stable symmetric solutions. They occur for the families without embedded vorticity having $l = 3$ and $l = 4$ at $G \approx 8.77$ and $G \approx 19.35$, respectively. A manifestation of this is the extended region of stability shown by the families $\Phi_{30}$ and $\Phi_{40}$ for $G > 0$ in the central panel Fig.~\ref{fig:m=0}.
\section{Charge-one vortices}
So far we have considered the nonlinear continuation of spherical harmonics without embedded vorticity. Now we proceed with the case $|m| = 1$ where the linear solutions are the spherical harmonics $Y_l^{\pm 1}(\theta,\phi)$ with $l = 1,2,\dots$. Since they have a position dependent phase $\exp(\pm i \varphi)$, we are now dealing with superfluid currents confined to the sphere. States that are nonzero in the vicinity of either pole display a vortex with its core exactly at the pole, thus compensating the diverging $\sin^{-2}(\theta)$ term in the Laplacian.

\begin{figure*}[htbp]
	\centering
    \includegraphics[width=453.17244pt,height=244.15715pt]{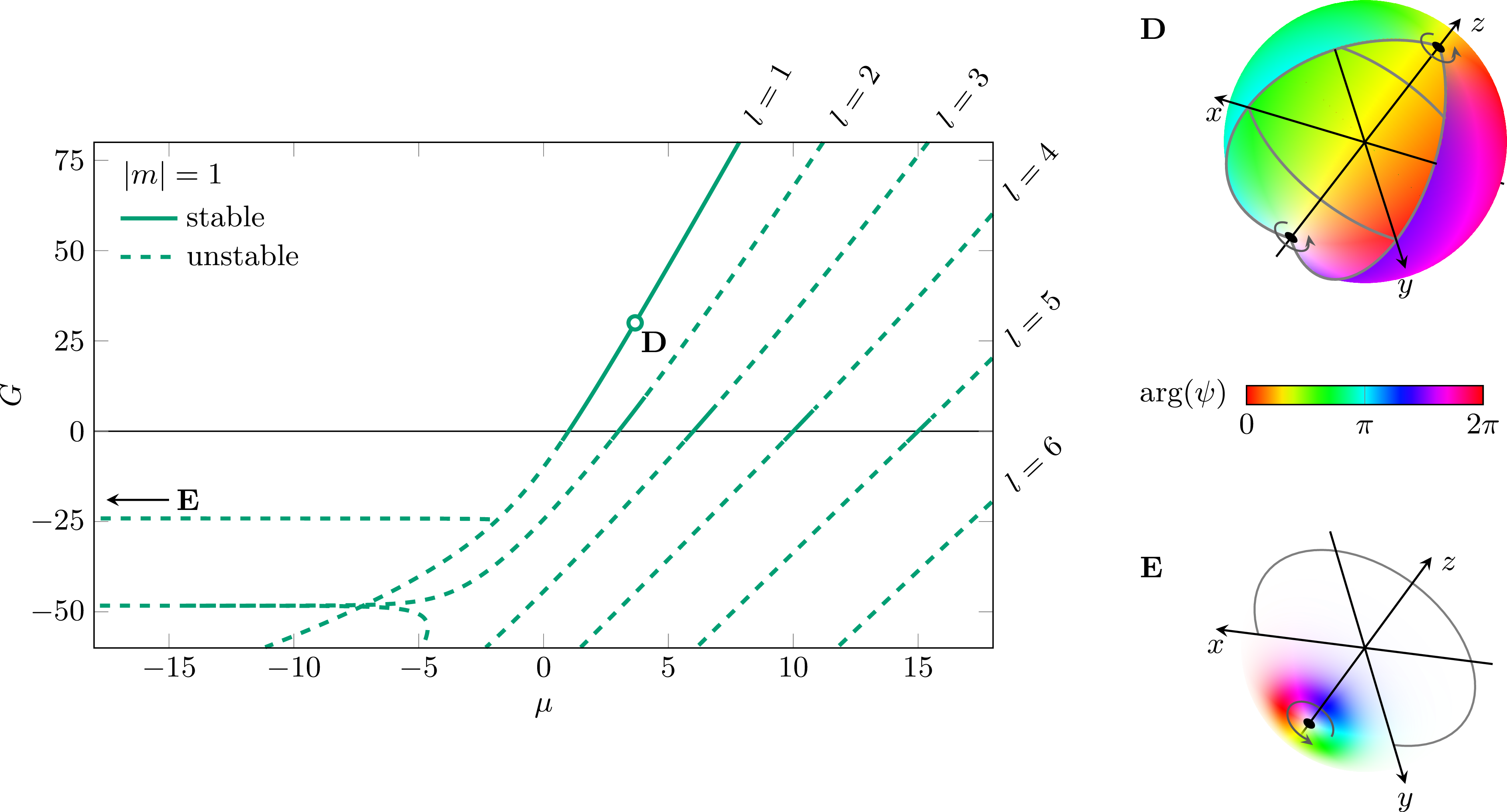}
    \caption{\label{fig:m=1}
        \textbf{Stationary states with embedded vorticity.}   
         Families of nonlinear modes on the $(\mu,G)$ diagram bifurcating form the linear limit, $G=0$, labeled by respective $l$ quantum number (attached outside the frame). Intervals of linearly stable and unstable solutions are shown by solid and dashed lines, respectively. The two states displayed on the right illustrate vortices on the sphere, supported either by a BEC with repulsive interparticle interactions (top) or by a vortex soliton under attractive interactions (bottom). The phases (arguments) of the corresponding order-parameters are encoded in color, while the circular arrows indicate the directions of the associated superfluid currents.    
    }
\end{figure*}

Figure~\ref{fig:m=1} shows that the linear solutions allow for analytic continuation to the nonlinear regime and share a finite region stability in the vicinity of $G = 0$ for both repulsive and attractive interactions.
The family $l = 1$, which features a vortex at one pole and an antivortex at the other one (Fig.~\ref{fig:m=1}, state \textbf{D}), is stable for all $G > 0$. Such vortex-antivortex pairs, as well as the related phenomenon of the formation of an Abrikosov lattice on a spherical surface, were investigated recently in Refs.~\cite{Padavic.2020,Bereta.2021,Caracanhas.2022,Tononi.2024,Li.2023,Saito.2023,White.2024}.

However, these prior studies focused predominantly on repulsive interparticle interactions ($G > 0$) or externally driven rotation, where topological constraints on $S^2$ requires zero net vorticity and vortex-antivortex pairing. The question of whether an isolated single vortex can exist on a static, non-rotating spherical shell remained unexplored within stationary bifurcation theory. 

Here, we demonstrate that switching to attractive interparticle interactions ($G < 0$) fundamentally alters this picture. Self-focusing nonlinearity allows for a spontaneous breaking of spatial symmetry, supporting a single localized vortex state without an antivortex partner even at zero external rotation. As shown in Fig.~\ref{fig:m=1} (state~\textbf{E}), this state belongs to the Townes vortex family for $|m|=1$, which bifurcates continuously from the linear $l = 1$ spherical harmonic at $G \approx -25$ and approaches the critical Townes norm $G_\text{TV} \approx -24.15$. 

We note that while a recent study~\cite{Zhao.2026} reported localized vortex states in attractive spherical BECs, their mechanism relies on external rotation to stabilize the $m=1$ state. In contrast, the nonlinear continuation of linear spherical harmonics, considered here, uncovers the stationary families in attractive ($G < 0$) non-rotating condensates. Although these pure cubic Townes vortices are unstable against critical collapse, we find that adding a repulsive quintic nonlinearity $\delta |\psi|^4 \psi$ ($\delta > 0$) stabilizes the single-vortex solitons on $S^2$ (see supplemental material).
\section{Superposition states}
\begin{figure*}[tbhp]
	\centering
    \includegraphics[width=446.42429pt,height=181.47516pt]{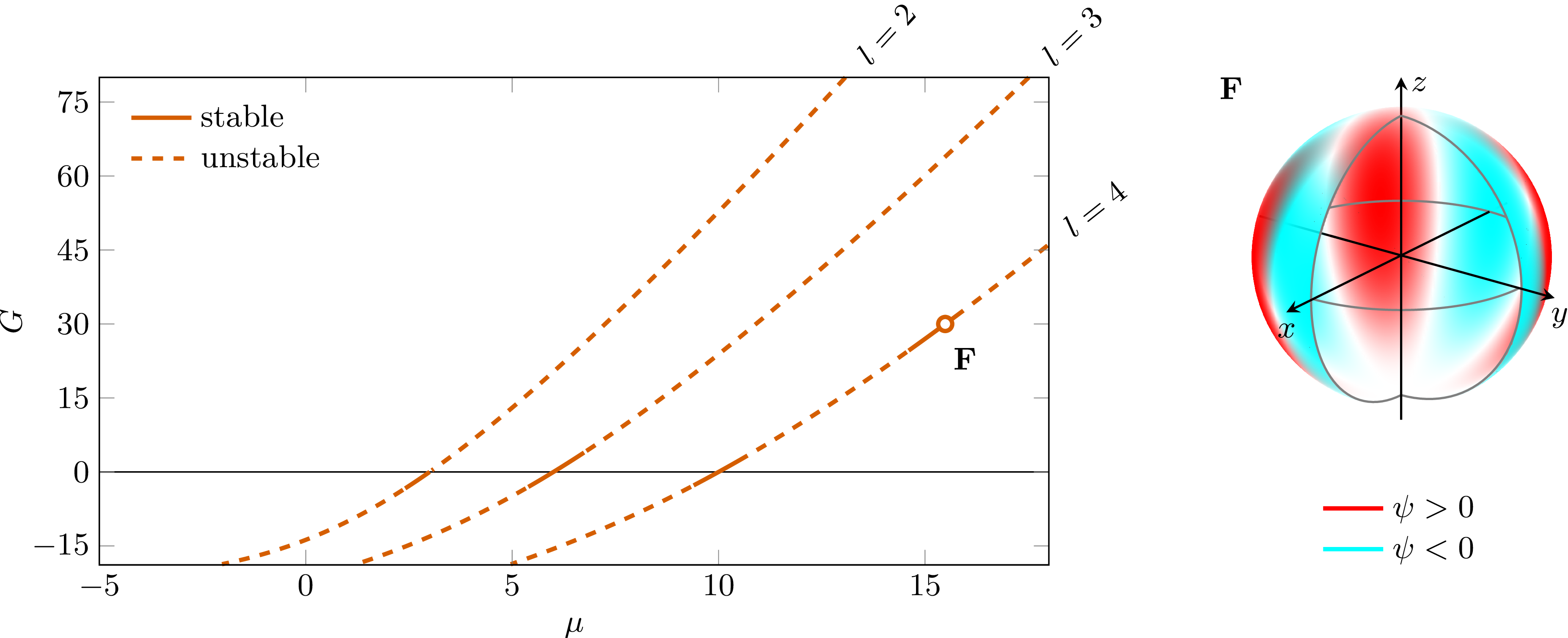}
    \caption{\label{fig:asymmetric}
        \textbf{Superposition states.} The combinations $Y_{l,l} + Y_{l,-l}$ are stationary in the linear regime $G = 0$ and can be extended into the nonlinear regime as shown in the $(\mu,G)$ diagram. Intervals of linearly stable and unstable solutions are displayed as solid and dashed lines, respectively. These states have completely asymmetric wavefunctions as exemplified on the right by the stable state ${\bf F}$.
    }
\end{figure*}

In the linear regime, $G = 0$, the stationary angular GPE~\eqref{eq:eigenvalue} has a degeneracy since $\tilde{\mu} = l (l + 1) / 2$ depends only on the $l$ number. Hence, for each $l$ there is a degenerate subspace with $2 l + 1$ solutions labeled by $m$~\cite{Landau.2003}. While any linear combinations of such states satisfy Eq.~\eqref{eq:eigenvalue} too, generally speaking, only certain combinations allow for analytical continuation to the nonlinear regime. An example are superpositions of $\pm m$ states which describe standing waves in $\varphi$-direction $\tilde{\Phi}_{lm}(\theta,\varphi) \propto P_l^m(\cos \theta) \cos(m \varphi)$. We restrict ourselves to the case $m = l$ where we show in Fig.~\ref{fig:asymmetric} that a continuation is indeed possible.
Note that there is no standing wave superposition for $l = 0$. For $l = 1$ it can be transformed to $\tilde{\Phi}_{10}(\theta^\prime,\varphi^\prime)$ by a proper rotation of the coordinate system $(\theta,\varphi) \to (\theta^\prime,\varphi^\prime)$.

As displayed in Fig.~\ref{fig:asymmetric}, the continuations are stable in the vicinity of $G = 0$ for $l = 2,3,4$ although only for $G < 0$ in the case of $l = 2$. Moreover, for $l = 4$ we observe a region of stability deeper into the repulsive nonlinear regime. The corresponding state features stripes at different longitudes as exemplified by state \textbf{F} in Fig.~\ref{fig:asymmetric}.
\section{Multi-mode breathing dynamics}
While unstable solutions (indicated by dashed lines in Figs.~\ref{fig:m=0}--\ref{fig:asymmetric}) decay and therefore cannot be observed directly, they may generate persistent dynamics. This is in particular the case for dark solitons which show snaking instability~\cite{Sallatti.2026} and decay into vortices. In the case of bifurcations we explore a similar scenario. Apart from unstable solutions one can find stable asymmetric ones with the same number of atoms, cf. Fig.~\ref{fig:m=0}. In such cases, the long-term dynamics observed within the spherical shell exhibits a notable property: an approximately periodic temporal evolution characterized by transitions between unstable and stable solutions and the participation of multiple modes. 

\begin{figure*}[tbhp]
	\centering
    \includegraphics[width=457.17883pt,height=299.61584pt]{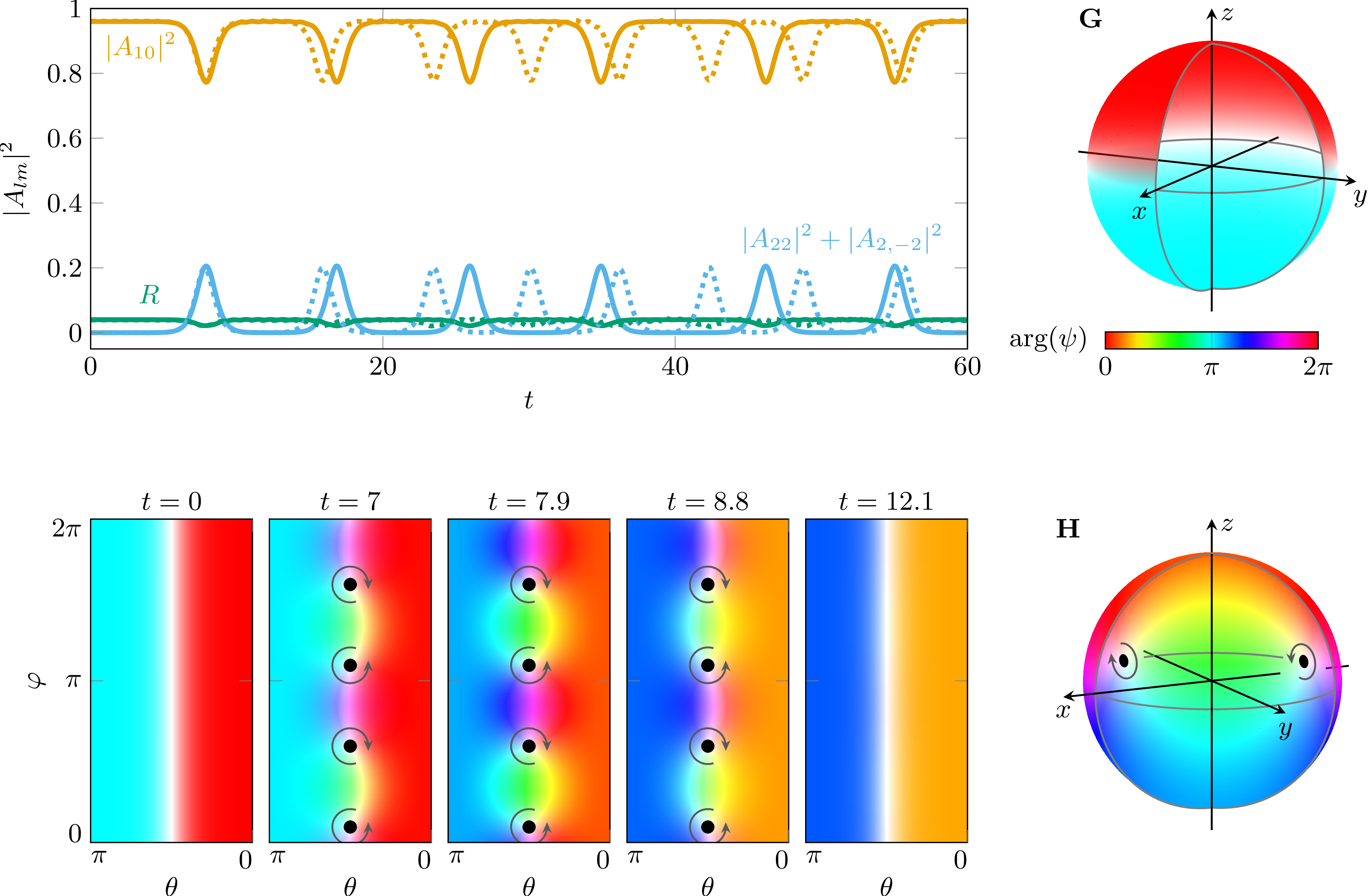}
    \caption{\label{fig:breather}
        \textbf{Multi-mode breathing dynamics.} Top panel: occupation of the dominating modes $|A_{10}|^2$, $|A_{22}|^2 + |A_{2,-2}|^2$ and residual occupation of all other modes $R = 1 - |A_{10}|^2 - |A_{22}|^2 - |A_{2,-2}|^2$ as a function of time are obtained within the Galerkin approximation (dotted) and high-resolution numerical simulation (solid). The dynamics describes a breathing between the initial dark soliton \textbf{G} and the four-vortex instant state \textbf{H} at $t = 7.9$. Location of these modes on the family $l = 1$ is shown in Fig.~\ref{fig:m=0}, where $\mathbf{H_\ast}$ is a stationary state close to \textbf{H}. Note that the sphere displaying solutions \textbf{G} and \textbf{H} are cut for better perspective. Therefore only two (of four) vortices, highlighted by curved arrows, are visible.
    }
\end{figure*}

To elucidate this scenario, we revisit the bifurcation of the $\Phi_{10}$ dark-soliton family discussed in Fig.~\ref{fig:m=0}, and choose an unstable state \textbf{G} at $G = 40$ (see Fig.~\ref{fig:m=0}) as an initial condition for numerical simulations of the GPE. The corresponding BdG spectrum exhibits a single pair of purely imaginary eigenfrequencies $(\omega,\omega^\ast)$, obtained for $|\mathfrak{m}|=2$. The obtained long-time evolution of the BEC is shown in Fig.~\ref{fig:breather}. The dark soliton initially undergoes a snaking instability previously reported in Ref.~\cite{Sallatti.2026}, which leads to its breakup into four vortices, as illustrated by the instant state \textbf{H} in Fig.~\ref{fig:breather}. This state is close to the stationary state $\mathbf{H_\ast}$ in Fig.~\ref{fig:m=0} with respect to number of atoms, symmetry, and number of vortices. The time evolution in Fig.~\ref{fig:breather} features dynamics with a few periods, i.e., the solution almost recovers its initial unstable state \textbf{G} at certain instances of time.

Such a few-mode dynamics stems from the discrete spectrum of the linear modes occurring due to finiteness of the surface sphere.  It is well described within, and thus is confirmed by, the Galerkin approximation accounting for only the most populated modes, which in the context of the vortex dynamics was previously used in Ref.~\cite{Driben.2016}. In its most elementary formulation, this reduces to the representation
\begin{equation}\label{Galerking}
    \psi \approx A_{10}(t) Y_{10} + A_{22}(t) Y_{22} + A_{2\,-2}(t) Y_{2\,-2} + R  
\end{equation}
where the mode with zero vorticity and two vortex modes with opposite charges $m=\pm 2$  are accounted, $A_{lm}(t)$ are the amplitudes of these modes, while $R$ stands for the rest of the expansion in terms of the spherical harmonics (see supplemental material for more details). Figure~\ref{fig:breather} compares the evolution of the amplitudes $A_{lm}$ obtained by projecting high-resolution simulations (solid) and the Galerkin approximation (dotted). 

Several salient features can be identified in the presented dynamics. First, it is observed that approximately 95\% of the atoms occupy only these three modes. Second, despite the instability of the state ${\bf G}$, the atomic population is found predominantly in this mode for most of the evolution (orange curve). Third, the trajectories exhibit a regular redistribution of population among the modes, although the exchange is not strictly periodic. Finally, while the Galerkin approximation (\ref{Galerking}) captures the qualitative characteristics of the evolution, quantitative discrepancies emerge at longer times: the intervals of enhanced intermodal population transfer are temporally shifted relative to those obtained from direct numerical simulations of the angular GPE.
\section{Conclusion}
Experimental verification of some solitonic solutions obtained in this article is feasible with available experimental setups. Townes solitons in a planar geometry have previously been created by suddenly quenching the interaction of a trapped BEC from the repulsive to the attractive regime~\cite{Chen.2020,Chen.2021}. The dynamics induced in this way featured trains of Townes solitons. In a similar fashion the homogeneous state on the shell can be quenched to the attractive regime which should result in the formation of bright solitons. This also allows to investigate the broken scale invariance because it should not be possible to rescale them all to a single profile anymore, as done in \cite{Chen.2020,Chen.2021} for the planar case.

The mixture-based shell~\cite{Jia.2022} is a perfect scheme for this proposal. Here $^{23}$Na atoms form a shell with radius $R \approx 10$~\textmu m around a core of $^{87}$Rb atoms. The time scale of this system is $\tau \approx 36$~ms which is sufficient to observe mean-field dynamics. By using a magnetic Feshbach resonance of $^{23}$Na at 907 Gauss~\cite{Chin.2010}, the intraspecies interaction can be tuned to realize the quench. At the same time the intraspecies $^{87}$Rb and the interspecies $^{23}$Na-$^{87}$Rb interaction stay repulsive~\cite{Chin.2010,Guo.2022} to keep the shell form alive.

On the International Space Station~\cite{Aveline.2020} ultracold atomic bubbles are created using a single atomic species in a radio-frequency dressed magnetic trap~\cite{Carollo.2022}, where it is challenging to tune interactions with magnetic Feshbach resonances. An alternative may be to release the atoms. After a short collimation step~\cite{Boegel.2023}, the freed-up magnetic fields could then be used to perform the quench. Since this setup allows for greater flexibility in choosing the radius of the shell, the time scale can be optimized to observe the formation of solitons.

Creating dark solitons on the sphere and observing their decay should work in full analogy to the planar case by using phase imprinting~\cite{Burger.1999,Denschlag.2000,Anderson.2001,Becker.2008}. Shining light on one half of the sphere to imprint a phase difference of $\pi$ produces state G in Fig.~\ref{fig:breather}. If this is done at a nonlinearity where the state is unstable, the multi-mode breathing dynamics can be experimentally probed.
\begin{acknowledgments}
	The work of VVK  was supported by the Fundação para a Ciência e Tecnologia under the project UID/00618/2025 (DOI https://doi.org/10.54499/UID/00618/2025). The authors gratefully acknowledge the scientific support and HPC resources provided by the German Aerospace Center (DLR). The HPC system CARA is partially funded by the Saxon State Ministry for Economic Affairs, Labour and Transport and the Federal Ministry for Economic Affairs and Climate Action. The authors acknowledge support by the state of Baden-Württemberg through bwHPC and the German Research Foundation (DFG) through Grant No. INST 40/575-1 FUGG (JUSTUS 2 cluster). This work was supported by the Science Sphere Quantum Science of Ulm University and the Center for Integrated Quantum Science and Technology (IQST). The research of the IQST is financially supported by the Ministry of Science, Research and Arts Baden-Württemberg.
\end{acknowledgments}
\section*{AUTHOR DECLARATIONS}
\subsection*{Conflict of interest}
The authors declare no competing interests.
\subsection*{Author contributions}
\textbf{A. Wolf:} Conceptualization; Formal analysis; Methodology; Software; Validation; Visualization; Writing – original draft.
\textbf{V. V. Konotop:} Conceptualization; Formal analysis; Methodology; Validation; Writing – review \& editing.
\textbf{M. A. Efremov:} Conceptualization; Methodology; Validation; Writing – review \& editing.
\appendix
\section*{Appendix}
To obtain the stationary states of the angular GPE, i.e. solve the nonlinear eigenvalue problem of Eq.~\eqref{eq:eigenvalue}, we use a new method developed in ~\cite{Altmann.2021}. The method relies on the standard shift-and-invert technique enhanced by a Rayleigh quotient iteration. The crucial step is the linearization of the right-hand side of Eq.~\eqref{eq:eigenvalue} which yields the iteration rule $[J(u^{n}) + \sigma ]u^{n + 1} = u^{n}$ with the operator
\begin{equation}\label{eq:jOperator}
    \begin{aligned}
        J(u) \Phi =& \left(\frac{1}{2} \bL^2 + G |u|^2\right) \Phi + G\bigg[u^2 \Phi^\ast + |u|^2 \Phi\\
        & - u|u|^2 \int \sin(\theta) \mathrm{d}\theta \mathrm{d}\varphi \ \left(u \Phi^\ast + u^\ast \Phi\right)\bigg]\,.
    \end{aligned}
\end{equation}
Given a good initial guess $u^0$ and shift $\sigma$, shift-and-invert quickly converges to the closest stationary state.

To represent the operator~\eqref{eq:jOperator} as matrix, we have to use a Nambu representation $(\Phi,\Phi^\ast)^\text{T}$, i.e. also consider the action of $J$ on $\Phi^\ast$. The squared angular momentum operator is implemented using spherical harmonic transformations. To solve the linear system in each iteration we use GMRES and a rank-1 update~\cite{Montoison.2023}. 

Linear stability of a stationary solution $\Phi(\theta,\varphi)$ is investigated by inserting the ansatz $\psi = \left(\Phi + u \mathrm{e}^{-\mathrm{i} \omega t} + v^\ast \mathrm{e}^{\mathrm{i} \omega t}\right) \mathrm{e}^{- \mathrm{i} \mu t}$ into Eq.~\eqref{eq:GPE}. The equation obtained in this way is linearized with respect to $u(\theta,\varphi)$ and $v(\theta,\varphi)$ which yields the Bogoliubov-de Gennes (BdG) equations
\begin{equation}\label{eq:bdgMatrix}
    \omega \begin{pmatrix}
        u \\ v
    \end{pmatrix} = \begin{pmatrix}
        \mathcal{L} & U\\
        -U^\ast & -\mathcal{L}
    \end{pmatrix} \begin{pmatrix}
        u \\ v
    \end{pmatrix}\,,
\end{equation}
where $\mathcal{L} = \frac{1}{2} \bL^2 + 2 G |\Phi|^2 - \mu$ and $U = G \Phi^2$. More details on the structure of the BdG spectrum are provided in the supplemental material.
\providecommand{\noopsort}[1]{}\providecommand{\singleletter}[1]{#1}%
\onecolumngrid
\clearpage
\setcounter{equation}{0}
\renewcommand{\theequation}{S.\arabic{equation}}
\setcounter{figure}{0}
\renewcommand{\thefigure}{S\arabic{figure}}
\setcounter{table}{0}
\renewcommand{\thetable}{S\arabic{table}}
\begin{center}
	{\normalfont\large\bfseries\centering {\it Supplemental material for}\\ 
		Bose-Einstein condensate in a shell: nonlinear modes and symmetry breaking}\\[23pt]	
\end{center}
\section*{Dimensional reduction}
Consider a 3D GPE equation
\begin{align}\label{GPE3D}
    \mathrm{i} \hbar \frac{\partial \Psi}{\partial t} = -\frac{\hbar^2}{2 M} \left(\frac{1}{r^2} \frac{\partial}{\partial r} r^2 \frac{\partial}{\partial r} - \frac{1}{r^2} \bL^2\right) \Psi + V(r) \Psi + g |\Psi|^2 \Psi
\end{align}
for the order parameter $\Psi = \Psi(r,\theta,\varphi,t)$, written down in spherical coordinates $(r,\theta,\varphi)$, and $\bL^2$ is the total angular momentum operator. The parameter $g = 4 \pi N \hbar^2 a / M$ quantifies the effect of atom-atom interaction, where $a$ and $N$ are the 3D $s$-wave scattering length and the total number of atoms, accordingly. The potential $V(r)$ significantly differs from zero in a narrow vicinity of the spherical surface having the radius $R$ (this condition is specified below).  

The weak nonlinearity allows us to neglect the nonlinear term in Eq.~\eqref{GPE3D} within the leading approximation. As a result, an approximate solution of Eq.~\eqref{GPE3D} can be searched in the factorized form
\begin{align}\label{ansatz_factor}
    \Psi = \exp\left(-\frac{\mathrm{i}}{\hbar} \varepsilon t\right) \frac{u(r)}{r} \psi(\theta, \varphi,t) \,,
\end{align}
where $u(r)$ is a normalized function,
\begin{align}\label{Normalization_u0}
    \int_0^\infty \mathrm{d} r \, u^2(r) = 1,
\end{align}
solving the eigenvalue problem
\begin{align}
    \label{eigen_r}
    -\frac{\hbar^2}{2 M} \frac{\mathrm{d}^2}{\mathrm{d} r^2} u + V(r) u = \varepsilon u, \quad r \in [0,\infty)
\end{align} 
with the boundary conditions $u(0) = 0$ and $\lim_{r \to \infty} u(r) = 0$~\cite{Landau.2003}.

Let $u_n(r)$ denote the eigenfunction of Eq.~\eqref{eigen_r} corresponding to the eigenvalue $\varepsilon_n$, with $n = 0$ corresponding to the ground state. Strong confinement of the atoms to a spherical surface is determined by the condition that the difference between the two lowest levels $\varepsilon_0$ and $\varepsilon_1$ is much larger compared to the energy of the nonlinear interactions, i.e., $(g/2) \int \mathrm{d} {\bf r} \, |\Psi|^4 \ll |\varepsilon_1 - \varepsilon_0|$. In this case, the consideration can be limited to the ground state only. Substituting $u(r)$ by the ground-state wave function $u_0(r)$ and $\varepsilon$ by $\varepsilon_0$ in Eqs.~\eqref{ansatz_factor} and~\eqref{GPE3D}, and integrating over $r$, one obtains the equation
\begin{align}\label{GPE3D_av}
    \mathrm{i} \hbar \frac{\partial \psi}{\partial t} = \frac{\hbar^2}{2 I} \bL^2 \psi + {\tilde g} |\psi|^2 \psi
\end{align}
describing the time dynamics of the angular wave function $\psi(\theta,\varphi,t)$. It slowly depends on time on the scale $\hbar / \varepsilon_0$. Here we have introduced the effective momentum of inertia
\begin{align}\label{Momentum_inertia_I}
   I = M \left(\int_{0}^\infty \mathrm{d} r \, \frac{u_0^2}{r^2}\right)^{-1}
\end{align}
and the effective nonlinear coefficient
\begin{align}
    \tilde{g} = g\int_{0}^\infty \mathrm{d} r \, \frac{u_0^4}{r^2}
\end{align}
in terms of the radial wave function $u_0(r)$. 

Introducing the dimensionless time $\hbar t / I$, one arrives at the dimensionless form of Eq.~\eqref{GPE3D_av}
\begin{align}\label{GPE3D_av_dimensionless}
    \mathrm{i} \frac{\partial \psi}{\partial t} = \frac{1}{2} \bL^2 \psi + G|\psi|^2 \psi
\end{align}
with the dimensionless parameter
\begin{align}
    G = \frac{I{\tilde g}}{\hbar^2} = 4 \pi a N \left(\int_{0}^\infty \mathrm{d} r \, \frac{u_0^4}{r^2}\right) \left(\int_{0}^\infty \mathrm{d} r \, \frac{u_0^2}{r^2}\right)^{-1}.
\end{align}

For $u_0(r)$ strongly localized around $r = R$, we obtain from Eqs.~\eqref{Normalization_u0} and~\eqref{Momentum_inertia_I} the result $I\approx MR^2$. As an example, for
\begin{align}
    u_0(r) = \frac{1}{\sqrt{\Delta R}}
    \begin{cases}
    1, & |r - R| \leq \Delta R / 2 \\
    0, & |r - R| > \Delta R / 2
    \end{cases}
\end{align}
with $R \gg \Delta R$, one derives $G = 4 \pi N (a / \Delta R)$ and $I = M \left(R^2 - \Delta R^2 / 4\right) \approx M R^2$.
\section*{Proof that \texorpdfstring{$\phi_{lm}(\theta)$}{the symmetric stationary state} is real}
\label{sec:reality}
Making the ansatz $\psi(\theta,\varphi,t) = \phi_{lm}(\theta) \mathrm{e}^{\mathrm{i} m \varphi - \mathrm{i} \mu t}$ in the GPE [Eq.~\eqref{eq:GPE} from the main text] we obtain
\begin{equation}\label{A1}
    \mu \phi_{lm} = -\frac{1}{2} \left(\Delta_\theta - \frac{m^2}{\sin^2(\theta)}\right) \phi_{lm} + G |\phi_{lm}|^2 \phi_{lm}, \quad \Delta_\theta \equiv \frac{\mathrm{d}^2}{\mathrm{d} \theta^2} + \cot(\theta) \frac{\mathrm{d}}{\mathrm{d} \theta}.
\end{equation}

Next, we follow the arguments of Ref.~\cite{Alfimov.2002}. Indeed, by allowing $\phi_{lm}$ to have complex values, we represent it in the exponential form
\begin{equation}
    \phi_{lm}(\theta) = \rho(\theta) \exp\left[\mathrm{i} \int_0^\theta \mathrm{d} \theta' \, \chi(\theta')\right]
\end{equation}
with real valued functions $\rho(\theta)$ and $\chi(\theta)$.

Now we substitute this expression into Eq.~\eqref{A1} and separating real and imaginary parts. The imaginary part is resolved readily with respect to $\chi(\theta)$ yielding
\begin{align}\label{A2}
    \chi(\theta) = \frac{C}{\sin(\theta) \left[\rho(\theta)\right]^2},
\end{align}
where $C$ is a real constant. 

With this result, the real part acquires the form
\begin{align}
    \mu \rho = -\frac{1}{2} \Delta_\theta \rho + \frac{m^2}{2 \sin^2(\theta)} \rho + G \rho^3 + \frac{C^2}{2 \rho^3 \sin^2(\theta)}.
\end{align}
Under constraint of boundness of $\rho(\theta)$, the last term in this equation diverges at $\theta = 0,\pi$, as well as at all points (if any) where $\rho = 0$. In such cases the equation does not have a solution, unless, however, $C = 0$. In this last case, from Eq.~\eqref{A2} we obtain $\chi \equiv 0$ and thus $\phi_{lm}(\theta) = \rho(\theta)$, i.e. $\phi_{lm}(\theta)$ is a real function.
\section*{Linear stability}
To investigate the linear stability of stationary states, we insert the ansatz
\begin{equation}\label{eq:bdgAnsatz}
    \psi = \left(\Phi + u \mathrm{e}^{-\mathrm{i} \omega t} + v^\ast \mathrm{e}^{\mathrm{i} \omega t}\right) \mathrm{e}^{- \mathrm{i} \mu t}
\end{equation}
into Eq.~\eqref{eq:GPE}. The small perturbations $u(\theta,\varphi)$, $v(\theta,\varphi)$, and corresponding exponents $\omega$ satisfy the Bogoliubov-de Gennes (BdG) equations
\begin{equation}\label{eq:bdgFull}
    \begin{aligned}
        \omega u &= \left(\frac{1}{2} \bL^2 + 2 G |\Phi|^2 - \mu \right) u + G \Phi^2 v\\
        -\omega v &= \left(\frac{1}{2} \bL^2 + 2 G |\Phi|^2 - \mu \right) v + G \left(\Phi^\ast\right)^2 u
    \end{aligned}
\end{equation}
which we solve numerically.

Because the wavefunction $\Phi(\theta,\varphi)$ of the stationary state appears in the BdG equations~\eqref{eq:bdgFull}, its form determines whether we must solve a full two-dimensional eigenvalue problem or can reduce it to a one-dimensional problem via separation of variables. In the article we begin by considering solutions of the kind $\Phi(\theta,\varphi) = \phi_{lm}(\theta) \mathrm{e}^{\mathrm{i} m \varphi}$, where $m$ is an integer and $\phi_{lm}(\theta)$ is real. For this case, separating the mode functions according to $u(\theta,\varphi) = u_{\mathfrak{l}\mathfrak{m}}(\theta) \mathrm{e}^{\mathrm{i} (\mathfrak{m} + m) \varphi}$ and $v(\theta,\varphi) = v_{\mathfrak{l}\mathfrak{m}}(\theta) \mathrm{e}^{\mathrm{i} (\mathfrak{m} - m) \varphi}$ with an integer $\mathfrak{m}$, yields the one-dimensional problem
\begin{equation}
    \begin{aligned}
        \omega u_{\mathfrak{l}\mathfrak{m}} &= \left\{-\frac{1}{2} \left[\Delta_\theta - \frac{(\mathfrak{m} + m)^2}{\sin^2(\theta)}\right] + 2 G \phi_{lm}^2 - \mu \right\} u_{\mathfrak{l}\mathfrak{m}} + G \phi_{lm}^2 v_{\mathfrak{l}\mathfrak{m}}\\
        -\omega v_{\mathfrak{l}\mathfrak{m}} &= \left\{-\frac{1}{2} \left[\Delta_\theta - \frac{(\mathfrak{m} - m)^2}{\sin^2(\theta)}\right] + 2 G \phi_{lm}^2 - \mu \right\} v_{\mathfrak{l}\mathfrak{m}} + G \phi_{lm}^2 u_{\mathfrak{l}\mathfrak{m}}\,.
    \end{aligned}
\end{equation}

The mode functions $u(\theta,\varphi)$ and $v(\theta,\varphi)$ can be sorted into three categories according to whether the Krein signature 
\begin{equation}
    \mathcal{K} = \int \sin(\theta) \mathrm{d}\theta \mathrm{d}\varphi \, \left(|u|^2 - |v|^2\right)
\end{equation}
is positive, negative, or zero. The former two cases are restricted to modes with nonzero real eigenvalues $\mathrm{Re}(\omega) \neq 0$, $\mathrm{Im}(\omega) = 0$ which correspond to quasiparticle and quasihole excitations, respectively. A signature of zero indicates that the eigenvector belongs either to a mode with complex eigenvalue $\mathrm{Im}(\omega) \neq 0$ or to a Goldstone mode with eigenvalue $\omega = 0$.

\begin{figure*}[tbhp]
	\centering
    \includegraphics[width=430.78857pt,height=323.49019pt]{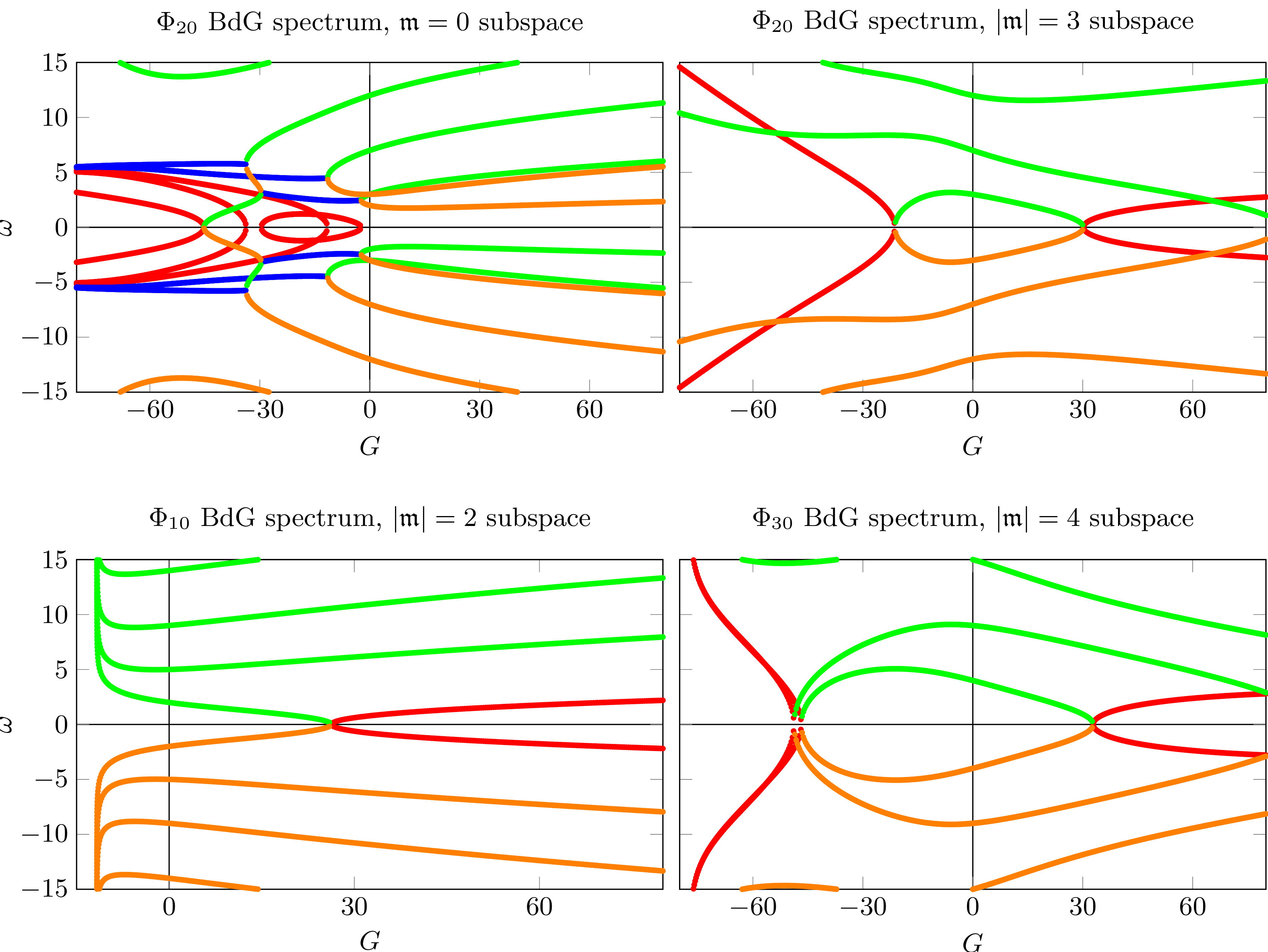}
    \caption{\label{fig:bdg}
        Bogoliubov-de Gennes spectra of selected nonlinear spherical harmonics with zero embedded vorticity $\Phi_{l0}$ at different nonlinearities $G$. Goldstone modes with $\omega = 0$ are excluded. The color indicates real (blue) and imaginary (red) part of complex eigenvalues or positive (green) and negative (orange) Krein signature of real eigenvalues.
    }
\end{figure*}

To give an example, we consider the stationary state of the angular GPE with $l = 2$, $m = 0$, where the BdG spectrum can be split into different $\mathfrak{m}$ subspaces, as discussed above. Figure~\ref{fig:bdg} (top left) shows the case of the $\mathfrak{m} = 0$ subspace and displays several characteristic features of the BdG equations.
Starting with $G > 0$ we observe that for every mode with eigenvalue $\omega$ and positive signature (green lines) there is one with eigenvalue $-\omega$ and negative signature (orange lines). This is due to the particle-hole symmetry of the BdG equations~\eqref{eq:bdgFull}:  $-\omega^\ast$, $(v^\ast,u^\ast)$ is a solution when $\omega$, $(u,v)$ is a solution.

In addition to the particle-hole symmetry, the BdG equations are also pseudo-Hermitian. When writing Eq.~\eqref{eq:bdgMatrix} as a matrix equation, the operator on the right hand side satisfies a relation of the form $M = \sigma_z M^\dagger \sigma_z$, where $\sigma_z$ is the Pauli matrix. Since $\sigma_z = \sigma_z^{-1}$, the BdG operator is similar to its Hermitian adjoint. Because similar matrices share their spectrum, when $\omega$ is an eigenvalue of the BdG equations, so is $\omega^\ast$. Combining the particle-hole symmetry with the pseudo-Hermicity shows that a complex eigenvalue always appears as quartet ($\omega$, $-\omega^\ast$, $\omega^\ast$, $-\omega$). When solving the BdG equations with an iterative eigensolver using shift-and-invert, it is therefore sufficient to scan the first quadrant of the complex plane.

Turning to attractive interaction $G < 0$ in Fig.~\ref{fig:bdg} (top left), we observe that two modes with opposite signatures (one green line and one orange line) may collide to a mode with zero signature. Two cases can be distinguished: (i) the modes collide at $\omega = 0$ and branch of into the complex plane with a purely imaginary eigenvalue (red lines emerge from the collision), or (ii) the modes collide at $\omega \neq 0$ and continue in the complex plane with finite real part (blue line emerges from the collision).

In scenario (i) an eigenvalue goes to zero and meets its own particle-hole partner. At the point of the collision the stationary state $\Phi(\theta,\varphi)$ either bifurcates or folds around in the $(\mu,G)$ diagram. We have used this as an indicator when searching for bifurcations. An example for a solution folding around is provided below. 

In scenario (ii), the stationary state $\Phi(\theta,\varphi)$ continuous as a unique state in the $(\mu,G)$ diagram without bifurcation, but develops an additional modulational instability. If the state was stable before the collision, because the whole BdG spectrum was real, it becomes unstable after the collision. Note that the particle-hole symmetry leads to a simultaneous collision at $-\omega$ which explains why the spectra in Fig.~\ref{fig:bdg} looks so symmetric with respect to the axis $\omega = 0$.

\section*{Two-dimensional plots for states \textbf{A}-\textbf{F}}
\begin{figure*}[htbp]
	\centering
    \includegraphics[width=451.73724pt,height=193.22417pt]{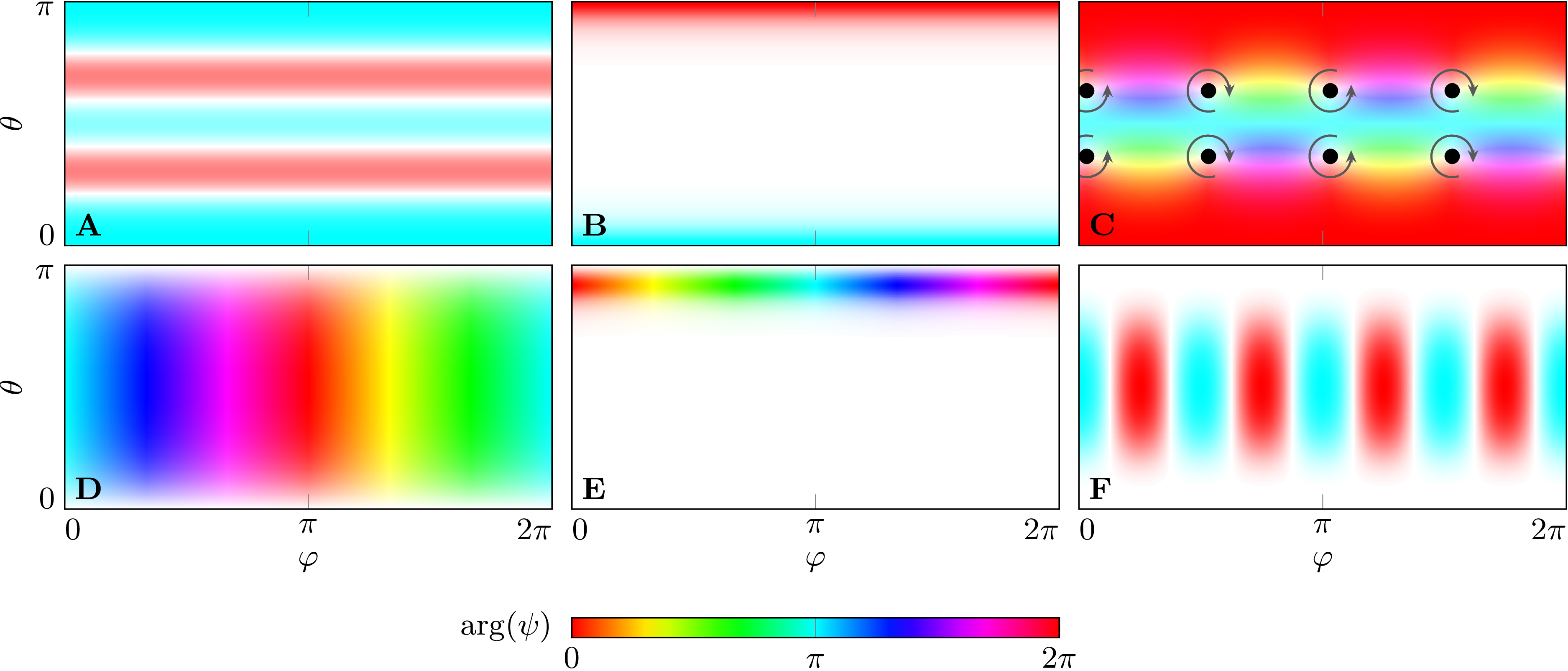}
    \caption{\label{fig:2dStates}
        Two-dimensional plots of states \textbf{A}-\textbf{F} presented in the main text. Color and opacity show phase and modulus of the wavefunction, respectively. 
    }
\end{figure*}

Figure~\ref{fig:2dStates} displays states \textbf{A}-\textbf{F} as two-dimensional plots to supplement the three-dimensional illustrations in the main text. Note that the top of each plot corresponds to $\theta = \pi$ and therefore the south pole (bottom) of the three-dimensional counterparts.
\section*{Stabilization of Townes-like solutions}
\begin{figure*}[tbhp]
	\centering
    \includegraphics[width=455.27014pt,height=133.02821pt]{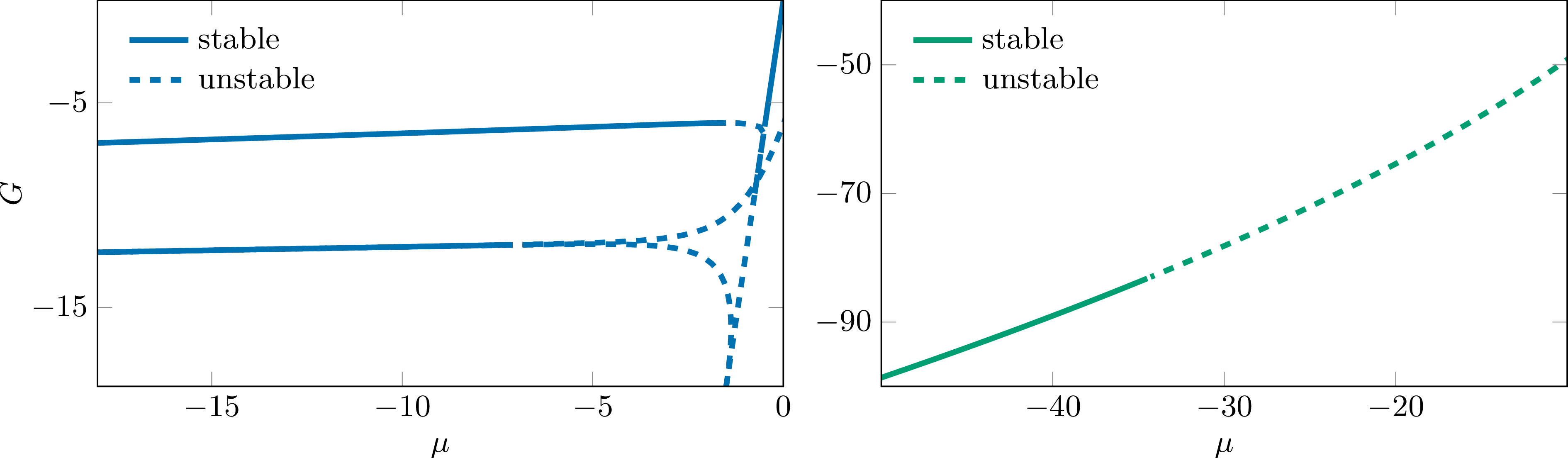}
    \caption{\label{fig:stabilization}
        Stabilization of Townes-like solitons using a higher order repulsive nonlinearity $\delta |\psi|^4$ with $\delta > 0$. Left panel shows the lower left quadrant of the $(\mu,G)$ diagram for $m = 0$, Fig.~\ref{fig:m=0} of the main text, using $\delta = 0.1$. Right panel shows stabilization of the single-vortex state with $m = 1$, as shown in Fig. \ref{fig:m=1} of the main text, using $\delta = 30$.
    }
\end{figure*}

Some Townes-like solutions can be stabilized by adding a higher order repulsive nonlinearity $\delta |\psi|^4$ with $\delta > 0$ into the angular GPE~\eqref{eq:GPE}. Very small values of $\delta$ (we have checked for $\delta = 0.01, 0.1,1$) are sufficient to stabilize a single Townes-like soliton or the out-of-phase combination, as depicted in Fig.~\ref{fig:stabilization} (left panel) for $\delta = 0.1$. Note that this term breaks the asymptotic scale invariance when the solutions are localized close to a point on the sphere. Consequently, the states display a downwards slope in the $(\mu,G)$ diagram, Fig.~\ref{fig:m=0}, instead of approaching a horizontal line. The solution with a single Townes-like soliton still bifurcates from the homogeneous state at $G \approx - 2 \pi$, as shown in Fig.~\ref{fig:m=0}, and only becomes stable after folding around in the $(\mu,G)$ diagram which coincides with the unstable BdG modes meeting at $\omega = 0$ and becoming real. 

The single vortex state, which is the lowest Townes-like soliton with $m = 1$, can be stabilized as well, albeit at much larger values of $\delta$. The right panel of Fig.~\ref{fig:stabilization} shows an example corresponding to $\delta = 30$.

\section*{Galerkin approach to multi-mode dynamics}
\label{sec:Galerkin}
The dynamics triggered by an unstable dark soliton, such as soliton \textbf{G} in Fig.~\ref{fig:breather}, can be described as an expansion in a series of the spherical harmonics
\begin{equation}
    \psi(\theta,\varphi,t) = \sum_{l=0}^\infty\sum_{m=-l}^l A_{lm}(t) Y_{lm}(\theta,\varphi)
\end{equation}
which form a complete orthonormal basis on the surface of sphere. 

Inserting the expansion into the angular GPE~\eqref{eq:GPE} yields the ordinary differential equation (ODE)
\begin{equation}\label{eq:ODE}
    \mathrm{i} \dot{A}_{l_1 m_1} = \frac{l_1 (l_1 + 1)}{2} A_{l_1 m_1} + G \sum_{l_2, m_2} \sum_{l_3, m_3} \sum_{l_4, m_4} A_{l_2,m_2}^\ast A_{l_3,m_3} A_{l_4,m_4} \int \sin(\theta) \mathrm{d}\theta \mathrm{d}\varphi \, Y_{l_1,m_1}^\ast Y_{l_2,m_2}^\ast Y_{l_3,m_3} Y_{l_4,m_4}
\end{equation}
for the coefficient \(A_{l_1 m_1}(t)\). 

The coupling between these coefficients induced by the nonlinear term is represented by the second term on the right-hand side, in which the integrals are evaluated explicitly as
\begin{equation}
    \begin{aligned}
        &\int \sin(\theta) \mathrm{d}\theta \mathrm{d}\varphi \, Y_{l_1,m_1}^\ast Y_{l_2,m_2}^\ast Y_{l_3,m_3} Y_{l_4,m_4} = \frac{\sqrt{(2 l_1 + 1)(2 l_2 + 1)(2 l_3 + 1)(2 l_4 + 1)}}{4 \pi} \times\\
        &\times \sum_{l = 0}^\infty \sum_{m = -l}^l (-1)^{m_1 + m_2 + m} (2l + 1)
        \begin{pmatrix}
            l_1 & l_2 & l \\
            0 & 0 & 0 
        \end{pmatrix}
        \begin{pmatrix}
            l_3 & l_4 & l \\
            0 & 0 & 0 
        \end{pmatrix}
        \begin{pmatrix}
            l_1 & l_2 & l \\
            -m_1 & -m_2 & m 
        \end{pmatrix}
        \begin{pmatrix}
            l_3 & l_4 & l \\
            m_3 & m_4 & -m 
        \end{pmatrix}
    \end{aligned}
\end{equation}
using $3j$ symbols~\cite{Varshalovich.1988}.

Setting $A_{lm}(t) = a_{lm} \mathrm{e}^{-\mathrm{i} \mu t}$ reduces the system~\eqref{eq:ODE} to a nonlinear root-finding problem. Note that the set of equations has to be constrained to the surface $\sum_{lm} |a_{lm}|^2 = 1$, to determine the chemical potential $\mu$ self-consistently alongside the coefficients $a_{lm}$. A valid root then corresponds to the stationary state
\begin{equation}\label{eq:stationaryGalerkin}
    \phi(\theta,\varphi) = \sum_{lm} a_{lm} Y_{lm}(\theta,\varphi)\,.
\end{equation}

The BdG equations can be dealt with in a similar fashion by inserting the stationary state~\eqref{eq:stationaryGalerkin} into Eqs.~\eqref{eq:bdgFull} and expanding
\begin{equation}\label{eq:bdgExpansions}
    u(\theta,\varphi) =  \sum_{\mathfrak{l}\mathfrak{m}} b_{\mathfrak{l}\mathfrak{m}} Y_{\mathfrak{l}\mathfrak{m}}(\theta,\varphi)\,, \quad v(\theta,\varphi) =  \sum_{\mathfrak{l}\mathfrak{m}} c_{\mathfrak{l}\mathfrak{m}} Y_{\mathfrak{l}\mathfrak{m}}(\theta,\varphi)
\end{equation}
which leads to integrals over four spherical harmonics as considered above. The coefficients $b_{\mathfrak{l}\mathfrak{m}}$ and $c_{\mathfrak{l}\mathfrak{m}}$ are obtained by solving the resulting eigenvalue problem. Note that large parts of the spectrum $\omega$ may be cut away when the expansions~\eqref{eq:bdgExpansions} are truncated to a few terms.

A qualitative description is obtained by truncating the series expansions to a finite number of terms that provide the dominant contribution, a procedure commonly referred to as the Galerkin approximation (see, e.g., \cite{Antman.2004}). To find the Galerkin (few-mode) approximation suitable to describe the dynamics presented in Fig.~\ref{fig:breather}, we require a Galerkin space that contains the dark soliton, unstable BdG eigenvector, and corresponding density modulation to a reasonable degree. The soliton consists of $m = 0$ modes with odd $l$. Taking the first three, that is $l = 1,3,5$, into account for the root-finding yields a chemical potential $\mu_\text{Galerkin} \approx 5.52942$ at $G = 40$. This is in good agreement with the numerical implementation of Ref.~\cite{Altmann.2021}, resulting in $\mu \approx 5.50887$. The relative error is $1 - \mu / \mu_\text{Galerkin} \approx 0.00371581$ and a comparison of the corresponding coefficients are listed in Tab~\ref{tab:galerkinStationary}.
\begin{table}[htbp]
    \centering
    \caption{\label{tab:galerkinStationary}
        Comparison of coefficients for the dark soliton.
    }
    \begin{tabular}{c|c|c|c}
        & $a_{10}$ & $a_{30}$ & $a_{50}$\\ 
        \hline
        Galerkin & $0.980272$ & $-0.192398$ & $0.0452784$\\
        \hline
        Numerics & $0.979725$ & $-0.194208$ & $0.0476755$
    \end{tabular}
\end{table}

The unstable BdG eigenvectors are in the subspaces with $|\mathfrak{m}| = 2$ and consist of modes with even $\mathfrak{l} \geq 2$. We focus on the subspace $\mathfrak{m} = 2$ and extend our Galerkin space by $\mathfrak{l} = 2,4,6$ (root-finding and numerics yields $a_{22} = a_{42} = a_{62} = 0$). The resulting spectrum contains the eigenvalue pair $\omega_\text{Galerkin} \approx \pm 1.15769 \mathrm{i}$. Comparison of this value to the one of our high resolution numerical simulation, giving rise to $\omega \approx \pm 1.15268 \mathrm{i}$, determines the relative error $1 - \omega / \omega_\text{Galerkin} \approx 0.00432705 $. Table~\ref{tab:galerkinBdg} lists the corresponding coefficients, with the normalization given by
$\int \sin(\theta) \mathrm{d}\theta \mathrm{d}\varphi \ (|u|^2 + |v|^2) = 1$.
\begin{table}[htbp]
    \centering
    \caption{\label{tab:galerkinBdg}
        Comparison of coefficients for the unstable BdG eigenvector.
    }
    \begin{tabular}{c|c|c|c|c|c|c}
        & $b_{10}$ & $b_{30}$ & $b_{50}$ & $b_{22}$ & $b_{42}$ & $b_{62}$\\ 
        \hline
        Galerkin & $0$ & $0$ & $0$ & $0.686564$ & $-0.151067 - 0.0588376 \mathrm{i}$ & $0.0416529 + 0.0247477 \mathrm{i}$\\
        \hline
        Numerics & $0$ & $0$ & $0$ & $0.687151$ & $-0.147766 - 0.0583797 \mathrm{i}$ & $0.0410455 + 0.0256756 \mathrm{i}$\\
        \multicolumn{7}{c}{}\\
        & $c_{10}$ & $c_{30}$ & $c_{50}$ & $c_{22}$ & $c_{42}$ & $c_{62}$\\ 
        \hline
        Galerkin & $0$ & $0$ & $0$ & $-0.233598 + 0.645602 \mathrm{i}$ & $-0.00392791 - 0.162073 \mathrm{i}$ & $0.00909909 + 0.047588 \mathrm{i}$\\
        \hline
        Numerics & $0$ & $0$ & $0$ & $-0.261625 + 0.635396 \mathrm{i}$ & $0.00227764 - 0.158864 \mathrm{i}$ & $0.0081141 + 0.0477298 \mathrm{i}$
    \end{tabular}
\end{table}

The instability is simulated by using an initial state where the unstable BdG eigenvector with a small amplitude $\eta = 10^{-4}$ is added to the dark soliton. Following Eq.~\eqref{eq:bdgAnsatz} we have $\psi(\theta,\varphi,0) = \Phi(\theta,\varphi) + \eta [u(\theta,\varphi) + v^\ast(\theta,\varphi)]$ which requires a representation of $v^\ast(\theta,\varphi)$ in our Galerkin space. Taking the conjugate of the expansion~\eqref{eq:bdgExpansions} yields $v^\ast(\theta,\varphi) = \sum_{\mathfrak{l}\mathfrak{m}} (-1)^\mathfrak{m} c_{\mathfrak{l}\mathfrak{m}}^\ast Y_{_{\mathfrak{l},-\mathfrak{m}}}$ which shows that we need the modes $\mathfrak{m} = -2$, $\mathfrak{l} = 2,4,6$. 

\begin{figure*}[tbhp]
	\centering
    \includegraphics[width=327.98048pt,height=136.58194pt]{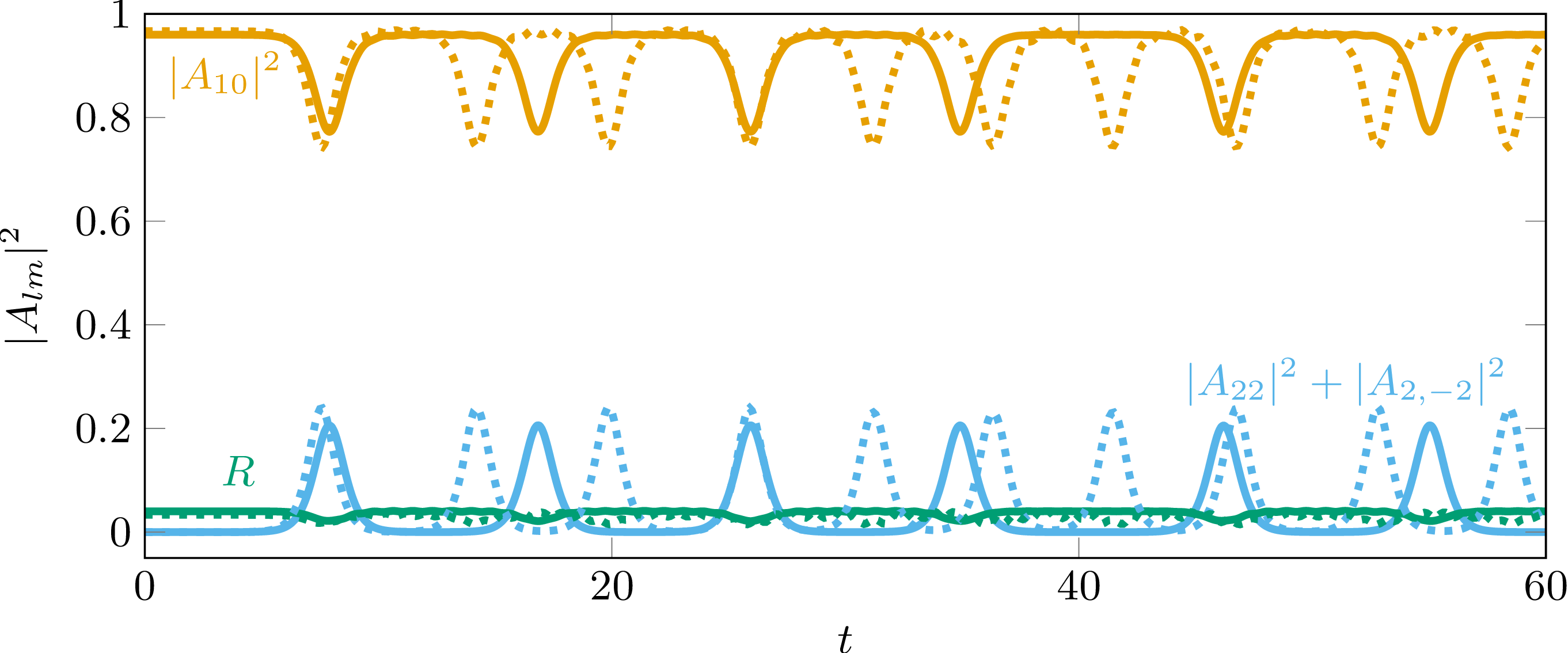}
    \caption{\label{fig:galerkin}
        Occupations $|A_{lm}|^2$ of selected spherical harmonics and residual occupation $R$ of all other modes as a function of time. In contrast to Fig.~\ref{fig:breather} the dotted lines use a reduced Galerkin space without the $l = 5$, $m = 0$ and $l = 6$, $m = \pm 2$ modes.
    }
\end{figure*}

In total we have nine spherical harmonics and the initial state for the ODE system is given in Tab.~\ref{tab:galerkinInitial}.
\begin{table}[htbp]
    \centering
    \caption{\label{tab:galerkinInitial}
        Initial condition for the ODE system to describe the multi-mode dynamics using the coefficients in Tabs.~\ref{tab:galerkinStationary} and~\ref{tab:galerkinBdg}.
    }
    \begin{tabular}{c|c|c|c|c|c|c|c|c}
        $A_{10}(0)$ & $A_{30}(0)$ & $A_{50}(0)$ & $A_{22}(0)$ & $A_{42}(0)$ & $A_{62}(0)$ & $A_{2,-2}(0)$ & $A_{4,-2}(0)$ & $A_{6,-2}(0)$\\
        \hline
        $a_{10}$ & $a_{30}$ & $a_{50}$ & $\eta b_{22}$ & $\eta b_{42}$ & $\eta b_{42}$ & $\eta c_{22}^\ast$ & $\eta c_{42}^\ast$ & $\eta c_{62}^\ast$
    \end{tabular}
\end{table}
The size of the Galerkin space is optimized to reproduce the first oscillation in Fig.~\ref{fig:breather} accurately. A purely qualitative understanding can already be gained without using the $l = 5$, $m = 0$ and $l = 6$, $m = \pm 2$ modes, as shown in Fig.~\ref{fig:galerkin}.
\end{document}